\documentclass[apj,iop, graphicx]{emulateapj}
\usepackage{apjfonts}

\newcommand\src{4U~1608$-$52}
\newcommand\xte{{\it RXTE}}
\newcommand\nustar{{\it NuSTAR}}

\makeatletter
\newcommand*{\rom}[1]{\expandafter\@slowromancap\romannumeral #1@}

\begin{document}

\title{Are the kHz QPO lags in neutron star 4U~1608$-$52 due to reverberation?}
\shortauthors{Cackett}
\shorttitle{Reverberation in 4U~1608$-$52?}

\author{Edward~M.~Cackett\altaffilmark{1}}

\email{ecackett@wayne.edu}

\affil{\altaffilmark{1}Department of Physics \& Astronomy, Wayne State University, 666 W. Hancock St, Detroit, MI 48201, USA}

\begin{abstract} 
X-ray reverberation lags have recently been discovered in both active galactic nuclei (AGN) and black hole X-ray binaries.  A recent study of the neutron star low-mass X-ray binary \src\ has also shown significant lags, whose properties hint at a reverberation origin.  Here, we adapt general relativistic ray tracing impulse response functions used to model X-ray reverberation in AGN for neutron star low-mass X-ray binaries.  Assuming relativistic reflection forms the broad iron line and associated reflection continuum, we use reflection fits to the energy spectrum along with the impulse response functions to calculate the expected lags as a function of energy over the range of observed kHz QPO frequencies in \src.  We find that the lag energy spectrum is expected to increase with increasing energy above 8 keV, while the observed lags in \src\ show the opposite behavior.   This demonstrates that the lags in the lower kHz QPO of \src\ are not solely due to reverberation. We do note, however, that the models appear to be more consistent with the much flatter lag energy spectrum observed in the upper kHz QPO of several neutron star low-mass X-ray binaries, suggesting that lower and upper kHz QPOs may have different origins. 
\end{abstract}

\section{Introduction}

Reflection of hard X-rays off an accretion disk has long been known to produce continuum and line emission, with the most prominent line being Fe~K$\alpha$ \citep[see][for a detailed review and history of X-ray reflection]{fabianross10}.  Reflection off the inner accretion disk around a black hole or neutron star will be subject to relativistic Doppler shifts and gravitational redshifts due to the high Keplerian velocities and deep gravitational potential there (\citealt{fabian89}, and see \citealt{miller07} for a review).  Such effects skew the line profile, with the amount of broadening depending on how close the accretion disk gets to the compact object.  Observations of broad, asymmetric profiles have been seen in Active Galactic Nuclei \citep[AGN, e.g.,][]{tanaka95,brenneman06, risaliti13}, stellar-mass black holes \citep[e.g.,][]{miller02, miller04, reis09_spin, miller15} and neutron star low-mass X-ray binaries \citep[LMXBs, e.g.,][]{bhattacharyya07, cackett08, cackett_j1808_09, cackett10, cackett12, pandel08, reis09_1705, disalvo09, disalvo15, iaria09, dai10, egron11, egron13, miller13, degenaar15}.

A natural prediction of X-ray reflection is that it should lead to a time lag between direct and reflected emission due to light travel time between the source of hard X-rays and the accretion disk \citep{stella90,campana95}.  The first significant X-ray reverberation lag was detected between the continuum and the soft excess in a long {\it XMM-Newton} observation of the AGN, 1H~0707$-$495 \citep{fabian09, zoghbi10, zoghbi11}, with the first Fe~K$\alpha$ lag detection discovered in NGC~4151 \citep{zoghbi12}.  Since then, many more X-ray reverberation lags have been seen \citep{emman11,demarco11,demarco13,cackett13, kara13a,kara13b,kara14, kara15,zoghbi13,zoghbi14,alston14,alston15, marinucci14}.  Evidence for thermal reverberation of the accretion disk has also been observed in stellar-mass X-ray binary GX 339$-$4 \citep{uttley11,demarco15} and H1743$-$22 \citep{demarco15}.  For a detailed review of X-ray reverberation, see \citet{uttley14}.

X-ray lags not associated with reverberation have been known of for a long time and have been observed in stellar-mass black holes  \citep[e.g.][]{page85,miyamoto88,miyamoto89,nowak99,uttley11}, AGN \citep[e.g.,][]{papadakis01, vaughan03, mchardy04, mchardy07}, and neutron stars \citep[e.g.][]{ford99}.    These are frequently referred to as `hard lags' since the hard X-ray continuum is seen to be lagging the soft X-ray continuum, and rather than being reverberation, are thought to be due to propagation of mass accretion rate variations though the corona \citep{kotov01,arevalo06,uttley11}.  In fact, both X-ray reverberation and hard lags are usually detected simultaneously but on different timescales, with hard lags detected at low frequencies and reverberation at high frequencies (see AGN X-ray reverberation references above, as well as \citealt{uttley11}).

The frequency and magnitude of X-ray reverberation lags in AGN scales with mass, as expected \citep{demarco13,kara13b}.  For example, for $M=10^7$~M$_\odot$, the soft lag is approximately 100 s at a frequency of $3\times10^{-4}$~Hz.  Scaling the mass to 1.4~M$_\odot$ for a neutron star, we would expect a lag of the order 14~$\mu$s at a frequency around 2000~Hz.  Of course, the geometry for an AGN and neutron star LMXB will be different, and the relative strength of the direct and reflection components will also differ \citep[see][for a discussion of dilution effects]{uttley14}.  For an alternative estimate, if we assume the inner accretion disk in a neutron star LMXB is not far from 6~GM/c$^2$ \citep[e.g., see][]{cackett10}, which for 1.4~M$_\odot$ is 12.4~km, and that the irradiating hard X-rays come from approximately the height of the neutron star $\sim$10 km, then the approximate distance is 16~km ($=\sqrt{10^2+12.4^2}$), which is a light travel time of 50~$\mu$s.   These estimates provide a guide as to the approximate lag amplitude and frequency where might expect to see reverberation in neutron star LMXBs.  

In order to detect a lag between two wavebands, the two lightcurves must have a high coherence \citep[see][for a discussion of coherence]{vaughannowak97}.  The highest frequency variability in neutron star LMXBs that match this requirement are kHz quasi-periodic oscillations \citep[QPOs; see][for a review]{vanderklis06}.  While the exact mechanism behind the formation of kHz QPOs is not fully understood, the high frequencies are close to what would be expected for Keplerian orbits at the inner disk \citep{miller98,stellavietri99}.  Lags in kHz QPOs are therefore a promising place to search for reverberation signatures.

Lags in kHz QPOs have been known for almost 20 years, and are usually soft lags \citep[e.g.][]{vaughan97,vaughan98,kaaret99}. While they have not been the focus of many studies, there has been a recent interest in kHz QPO lags once more \citep{barret13, deavellar13, peille15}.  Of particular interest is the hint of structure around the iron line region in the lag-energy spectrum of \src, which \citet{barret13} suggest could be due to reverberation -- directly converting the lag between the iron line and 15 keV to a size scale implies an inner disk radius of $\sim$9 GM/c$^2$.

The lags in \src\ have motivated this current study.  X-ray reverberation in AGN has led to the development of general relativistic ray-tracing impulse responses \citep{reynolds99,wilkins13, cackett14, emman14, chainakun15}.  The impulse response describes the response of the accretion disk to a delta-function flare, and encodes the geometry, kinematics and relativistic effects present there.  Here, we adapt these models for use in neutron star LMXBs, and determine the expected reverberation lags based on the energy spectrum of \src, and the observed frequencies of the kHz QPOs.  In section \ref{sec:data} we describe the data reduction and analysis, in section \ref{sec:model} we describe the model and apply it to the observed lags in \src.  Finally, in section~\ref{sec:disc} we discuss the implications of our results.

\section{Data Reduction and Analysis}\label{sec:data}

\subsection{Timing Analysis}

We analyze the {\it RXTE} ObsID 10072-05-01-00 observation of \src, following the same procedure as \citet{barret13}.  We choose this observation since it was performed with 64 spectral channels (higher than other observations), and showed potential evidence for a reverberation signature.

Following \citet{barret13}, we start by determining the power density spectrum (PDS) in the 3 -- 30 keV range for 128s intervals by averaging the PDS  using 32 segments of 4s.  For each 128s interval, we fit the PDS with a constant plus Lorentzian to determine the QPO frequency, $\nu_{\rm QPO}$.  We then shift and add all the PDS to the mean QPO frequency and determine the mean width, $\bar{w} = 5.2\pm0.1$~Hz.

\begin{figure}
\centering
\includegraphics[width=\columnwidth]{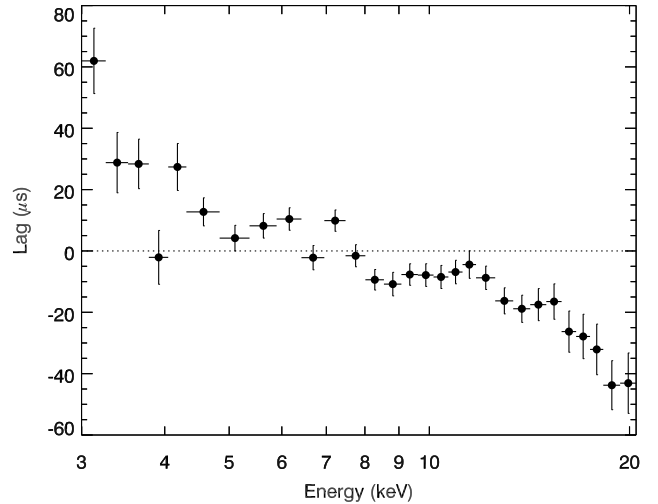}
\caption{QPO lag energy spectrum of \src\ from ObsID 10072-05-01-00.  The lags are calculated with respect to the 3 -- 30 keV reference band, with a positive lag indicating a lag behind the reference band.}
\label{fig:lagen}
\end{figure}

To determine the lag-energy spectrum, we compute the cross spectrum between the reference band (3 -- 30 keV) lightcurve and each channel, in turn, ensuring that the channel of interest is removed from the reference band \citep[this prevents correlated errors;][]{uttley11, uttley14}.  To determine the QPO lag, we average the cross spectrum over the frequency range $\nu_{\rm QPO} - \bar{w} \leq \nu_{\rm QPO} \leq \nu_{\rm QPO} + \bar{w}$ for each 128s segment, and calculate the time lag from the phase of the average cross spectrum.  We then take the weighted mean of the lags in that channel of interest over the full ObsID.  The resulting lag-energy spectrum we obtain is consistent with that of \citet{barret13}, and is shown in Figure~\ref{fig:lagen}.

\subsection{Spectral Analysis}

To build the most relevant transfer function including the full reflection spectrum and not just the Fe K$\alpha$ line, we fit the {\it RXTE}/PCA spectrum of \src\ from the same ObsID 10072-05-01-00.  We extract the Standard 2 spectrum from all available PCUs, adding a 0.6\% systematic error to each channel.  We fit the spectrum using \verb|xspec| v 12.8.2 \citep{arnaud96} in the 3 -- 20 keV energy range (the background begins to dominate above 20 keV).  

The X-ray spectrum of \src\ has been well studied previously \citep[e.g.,][]{gierlinski02,barret13,degenaar15}.  For this specific observation, \citet{barret13} has already shown that the spectrum is well fit by a disk blackbody, Comptonization component and a significant broad iron line, although continuum spectra of neutron star LMXBs are well known to be fit equally well with different models \citep[e.g.,][]{lin07}.  We too get good fits with such a model. However, if the broad iron line is formed due to reflection of hard X-rays off the accretion disk, then the iron line {\it must} be accompanied by the associated reflection continuum \citep[e.g.,][]{guilbert88,lightman88,george91}.  Since the purpose of this investigation is to explore whether the observed lags can be associated with reverberation, our spectral model needs to characterize the full reflection continuum, not just the iron line.  

\begin{figure}
\centering
\includegraphics[angle=270, width=0.9\columnwidth]{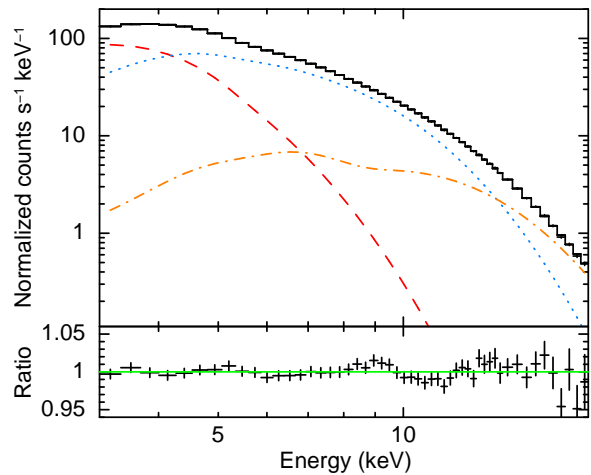}
\caption{{\it Top:} \xte/PCA spectrum of \src, with best-fitting model (black). The red dashed line shows the disk blackbody, the blue dotted line the single-temperature blackbody and the orange dash-dotted line the reflection component. {\it Bottom:} The ratio of the data to the best-fitting model.}
\label{fig:spec}
\end{figure}

For neutron star LMXBs in soft states, it is possible that the boundary layer between the neutron star and the accretion disk is the source of the flux that irradiates the disk, leading to reflection \citep[e.g.,][]{cackett10,dai10}.  In soft states the boundary layer is typically modeled by a Comptonized component, with a high optical depth, and hence can be well approximated by a blackbody.  We therefore fit a model consisting of an absorbed disk blackbody plus single temperature blackbody and relativistically blurred reflection model.  For the reflection model we use \verb|bbrefl|, a model for reflection of a blackbody off a constant density ionized disk \citep{ballantyne04} convolved with the \verb|rdblur| model for relativistic effects around a non-rotating black hole \citep{fabian89}.  The temperature of the irradiating blackbody in the reflection model was tied to the single temperature blackbody component.  Given the low spectral resolution of \xte/PCA, the relativistic parameters are not well constrained, thus, we use recent results from {\it NuSTAR} observations of \src\ as a guide.  Recently {\it NuSTAR} observed \src\  in a hard state, finding a strong Fe K$\alpha$ line, with the inner disk radius constrained to be 7 -- 10~$GM/c^2$ and an inclination around $40^\circ$ \citep{degenaar15}.  We therefore fix the inner disk radius at $8.5~GM/c^2$, and the inclination at $40^\circ$ in our spectral fits, furthermore, we fix the emissivity index at $\beta=-3$ and the outer radius of the accretion disk at 1000 $GM/c^2$.  We find that if we leave both the emissivity index, $\beta$, and the inner disk radius, $R_{\rm in}$, as free parameters in our fits, they are not well constrained.  The emissivity index unphysically pegs at the limit of $-10$, hence we make the standard choice of $\beta = -3$.  With $\beta=-3$, we obtain a best-fitting inner disk radius of 7.7~GM/c$^2$, but the 1$\sigma$ range is 6 -- 12.4~GM/c$^2$ (6 GM/c$^2$ is the lowest allowed limit to the inner disk radius), this is consistent with the {\it NuSTAR} measurements and our assumed value.
As the lower energy range of the fits is 3 keV, we choose to fix the Galactic photoelectric absorption (\verb|phabs|) at $N_{\rm H} = 1.5\times10^{22}$ cm$^{-2}$ \citep{penninx89}.  The best-fitting model and spectrum are shown in Figure~\ref{fig:spec} and the spectral parameters are given in Table~\ref{tab:spec}.  All  uncertainties are quoted at the 1$\sigma$ level.

We note that this full reflection model fits the spectrum only marginally better than a phenomenological fit (for a model consisting of diskbb+comptt+diskline we get $\chi^2 = 52.44$ for 44 degrees of freedom, compared to $\chi^2 = 48.44$ for 46 degrees of freedom for the reflection model presented here). However, we stress again that a broad iron line produced by reflection from the disk, must also be accompanied by the associated reflection continuum.  The \verb|bbrefl| model also assumes a single-temperature blackbody, and while this is a good approximation of a high optical depth Comptonized spectrum (we find only a 5\% difference in flux over 1 -- 20 keV between the best fitting Comptonized component, and a single-temperature blackbody), it might not accurately represent the spectrum of the irradiating flux.  We therefore also tried several other reflection models.  Reflection assuming an illuminating power-law, as is usually used for black hole sources, or hard states in neutron star low-mass X-ray binaries, cannot fit this spectrum of \src\ above 10 keV.  We also tried using the \verb|rfxconv| convolution model \citep{done06,kolehmainen11}, which allows the use of a Comptonized model as the irradiating flux, however, we did not find a stable fit \citep[as was also discovered in][]{barret13}.  We are therefore left with \verb|bbrefl| as our best-fitting full reflection model. We will discuss the implications of this on our results in Section~\ref{sec:disc}.

\begin{deluxetable}{ll}
\tablecaption{Spectral fit parameters}
\tablehead{\colhead{Parameter} & \colhead{Value}}
\startdata
$N_{\rm H}$ ($10^{22}$ cm$^{-2}$) & 1.5 (fixed)\\
$kT$ disk (keV) & $0.88\pm0.02$ \\
Disk normalization & $654\pm62$\\
$kT$ blackbody (keV) & $1.68\pm0.01$\\
Blackbody normalization & $(4.83\pm0.07)\times10^{-2}$\\
Emissivity & $-3$ (fixed) \\
$R_{\rm in}$ ($GM/c^2$) & 8.5 (fixed) \\
$R_{\rm out}$ ($GM/c^2$) & 1000 (fixed) \\
$i$ ($^\circ$) & 40 (fixed)\\
$\log \xi$ & $2.12\pm0.03$\\
Reflection normalization & $(4.4\pm0.3)\times10^{-25}$\\
3 -- 20 keV flux (erg s$^{-1}$ cm$^{-2}$) & $(5.5\pm0.1)\times10^{-9}$ \\
3 -- 20 keV unabs. disk flux (erg s$^{-1}$ cm$^{-2}$) & $(1.66\pm0.04)\times10^{-9}$\\
3 -- 20 keV unabs. bbody flux (erg s$^{-1}$ cm$^{-2}$) & $(3.47\pm0.04)\times10^{-9}$\\
3 -- 20 keV unabs. reflection flux (erg s$^{-1}$ cm$^{-2}$) & $(7.67\pm0.45)\times10^{-10}$\\
$\chi_\nu^2$ (dof) & 1.05 (46) 
\enddata
\label{tab:spec}
\end{deluxetable}

\section{Modeling the lags}\label{sec:model}

\begin{figure*}
\centering
\includegraphics[width=12cm]{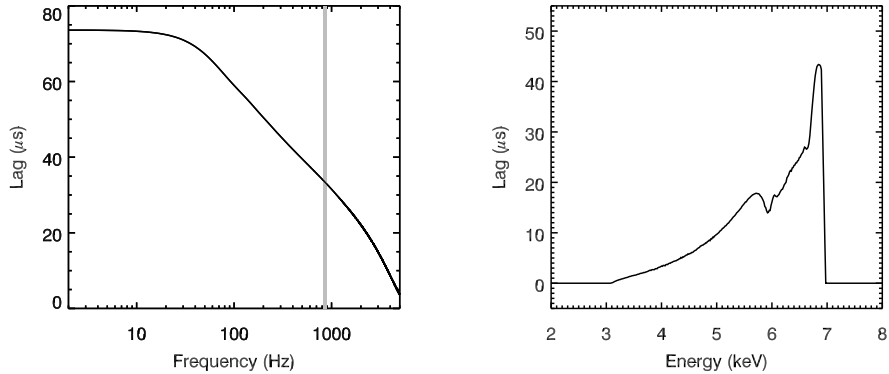}
\caption{{\it Left:} The average lag of the iron line as a function of frequency for $a = 0.29$, $h = 5~GM/c^2$ and $i=40^\circ$.  The gray shaded area indicates the frequency range of the kHz QPOs used here.  {\it Right:} The iron line lag averaged over the frequency range of the kHz QPOs (820 - 890 Hz).}
\label{fig:lagspec}
\end{figure*}

To start, we investigate the expected iron line lag at the frequency of the kHz QPOs in \src.  We do this by using the general relativistic ray tracing transfer functions of \citet{reynolds99}, as described by \citet{cackett14}, for a lamppost geometry (a point source above the compact object).  The parameters for the model require the height of the point source, spin of the compact object and inclination.  The spin frequency of \src\ is $\nu = 620$ Hz \citep{muno01}, which corresponds to a dimensionless spin parameter of $a=0.29$.  For $a=0.29$, the innermost stable circular orbit is at 5.0 $GM/c^2$, thus we choose, $h=5~GM/c^2$. Fitting of relativistic reflection models to the \nustar\ spectrum of \src, gives an inclination of approximately $40^\circ$ \citep{degenaar15}, which we adopt for our model.  Using these parameters we calculate the impulse response function.  Following \citet{cackett14}, we determine the frequency-dependence of the average iron line lag (left panel of Figure~\ref{fig:lagspec}) and the energy-dependence of the iron line lag at the frequency of the kHz QPOs (820--890 kHz, right panel of Figure~\ref{fig:lagspec}), assuming a reflected response fraction of 0.5, and a neutron star mass of 1.4 M$_\odot$.  From these models, we see that at the frequency of the kHz QPOs, there should indeed be an iron line lag of the order of around a few tens of $\mu$s, depending on the reflected response fraction.

The lags observed in \src, however, are not just at the iron line and are seen over the full energy range.  We therefore need to consider lags from the full reflection spectrum, rather than just the iron line.  Since both the reference band and the energy band of interest always contain both the direct and reflected components, the measured lag depends on the fraction of the variable flux in each component at a given energy -- the more flux from the direct component in the band of interest, the more the reflected lag gets diluted \citep[see, e.g.,][for a detailed discussion of dilution effects]{kara13_1h0707}.  As detailed in \citet{zoghbi13}, because of the dilution effects, in a given frequency band the energy-dependent lags trace the reflection fraction.  Determining the reflection fraction at all energies is therefore important.

\begin{figure}
\centering
\includegraphics[width=0.8\columnwidth]{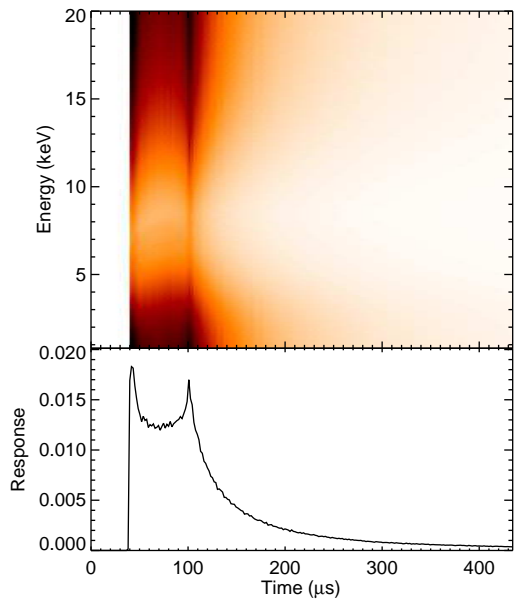}
\caption{{\it Top:} 2D impulse response function for reflection from an accretion disk with a source a height $h=5$ GM/c$^2$, spin $a = 0.29$, mass $M = 1.4$ M$_\odot$, and inclination $i=40^\circ$.  The reflection model has the parameters given in Table~\ref{tab:spec}, and the impulse response function is adjusted for the reflection fraction at each energy, as determined from spectral fitting.  {\it Bottom:} The energy-averaged response of the accretion disk. }
\label{fig:tf}
\end{figure}

Reflection off an accretion disk around a neutron star (or stellar-mass black hole) will lead to thermal emission from the accretion disk itself \citep[see, e.g.][]{ross07}, which is not included in the reflection model used here.  In fact, in stellar-mass black holes thermal reverberation from the disk has been detected \citep{uttley11,demarco15}.  However, our spectral modeling does include a disk blackbody component, thus, we use the shape of that emission to approximate the reflected thermal disk emission.  To determine the reflection fraction at a given energy we therefore include both the flux in the unblurred reflection component plus the disk blackbody component divided by the total flux at that energy, from our best-fitting spectral model to \src.  We then convolve this reflection fraction with the impulse response function -- this performs the relativistic blurring, as well as adding the time dimension.  This gives us a dilution-corrected impulse response function for the full reflection plus disk spectrum, which we show in Figure~\ref{fig:tf}.  We normalize the impulse response function by a reflected response fraction, defined at 6.4 keV.

\begin{figure}
\centering
\includegraphics[width=\columnwidth]{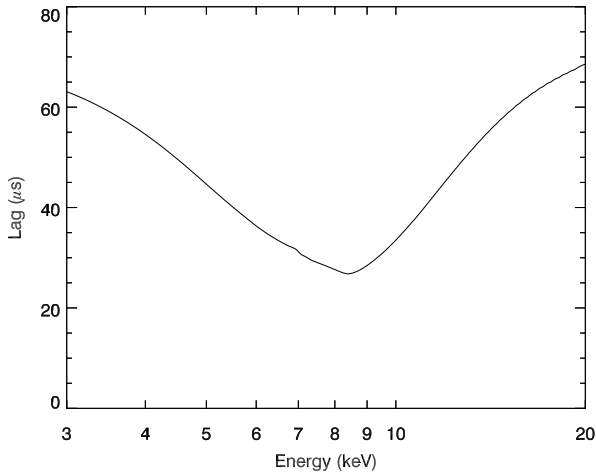}
\caption{Model lag-energy spectrum averaged over the frequency range 820--890 Hz, based on the observed energy spectrum. The trend of the lag increasing with energy above 8 keV is the opposite of what is observed in \src.}
\label{fig:lagmodel}
\end{figure}

We use the dilution-corrected impulse response function to determine the expected lag (as a function of energy) from reflection over the observed kHz QPO range (820--890 Hz), following the procedure described in \citet{cackett14}, and assuming a reflected response fraction of 0.5.  The resulting lags are shown in Figure~\ref{fig:lagmodel}.  Below 8 keV the lags decrease with increasing energy, while above 8 keV the lags increase with increasing energy.  Below 8 keV thermal emission from the disk is dominating the lags, and since the disk emission has an effective temperature of about 0.9 keV, the lags decrease with increasing energy, following the flux from the disk blackbody spectrum.  Above 8 keV reflection from the Compton hump becomes dominant.  Here, the reflection fraction increases steadily (as can be seen by comparing the model spectral components in Figure~\ref{fig:spec}), which translates into a larger lag at higher energies.  It is also important to note that the prominence of the Fe K line in the lags is significantly reduced when including lags from the full reflection spectrum, and a prominent Fe K lag is not predicted for the reflection spectrum in \src.

\begin{figure}
\centering
\includegraphics[width=\columnwidth]{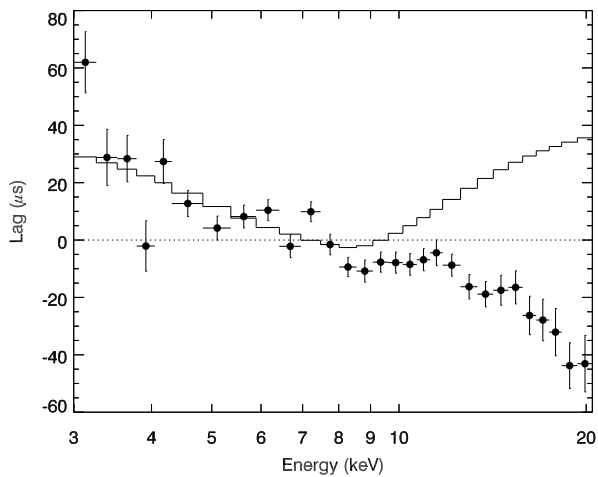}
\caption{Lag-energy spectrum of \src\ fit with the reflection model lags below 8 keV.  While the model matches the data well below 8 keV, it diverges significantly above 8 keV.}
\label{fig:modelfit}
\end{figure}

This overall shape is at odds with the observed lag spectrum of \src, which simply decreases with energy -- no upturn above 8 keV is observed.  To investigate in more detail we fit this model to the observed lags in \src\, allowing the reflected response fraction to be a free parameter.  Since the shape above 8 keV is clearly not consistent with the data, we only fit the model below 8 keV.  Remember that all the measured lags are relative, and so we are free to shift the model up or down by a constant value.  The best-fitting model is shown in Figure~\ref{fig:modelfit}, and requires a reflected response fraction of 0.37. While the model follows the observed lags well below 8 keV, it diverges significantly above 8 keV, clearly demonstrating that reflection alone cannot explain the lags in the lower kHz QPOs in \src.

\section{Discussion}\label{sec:disc}

Time lags observed in the lower kHz QPOs in the neutron star low-mass X-ray binary \src\ are of the right magnitude expected for reverberation off the accretion disk \citep{barret13}, and even hint at some structure around the Fe K line.  Here, we use general relativistic ray tracing models of the impulse response function, along with modeling the energy spectrum of \src, to develop a model for the lags (as a function of energy) expected from reverberation.

The exact shape of the lag energy spectrum depends strongly on dilution effects, i.e. what fraction of the emission at a given energy is from reflection, and what fraction is direct emission.  The lag energy spectrum thus traces the reflection fraction \citep[as described in detail by][]{zoghbi13}.  In the observations of \src\ used here, the source was in a soft state, with the energy spectrum falling off quickly above 10 keV, and having no significant hard power-law component.  The effect of this is that the reflection fraction increases significantly above 8 keV, and thus the lags are expected to increase with increasing energy above this value.  Below 8 keV reflected thermal emission from the accretion disk becomes dominant.  As the disk is most prominent at the lowest energies (since it has an effective temperature of 0.9 keV in this observation), we therefore expect the lags to decrease from 3 keV to 8 keV.  Such a decrease is observed, and can be fit by the reflection model we use, however, that model predicts an increase in the lags above 8 keV while the observed lags show the opposite behavior and continue with a near monotonic decrease with energy.

This result is dependent on the spectral deconvolution used, since this leads to the reflection fraction needed to predict the lags. Spectral modeling of neutron star LMXBs is notoriously degenerate, thus, it is important to assess the assumptions we make here.  Our basic assumption is that the broad iron line is formed due to relativistic reflection from the inner accretion disk.  Under such an assumption the spectral model must fit the full reflection spectrum, including iron line emission as well as the associated reflection continuum. We achieved a good spectral fit using a reflection model where the irradiating flux is a blackbody, as might be expected if the boundary layer irradiates the disk in the soft state. However, the true irradiating flux may take a slightly different spectral shape, for instance, a Comptonized component. Regardless of the exact spectral shape of the irradiating flux, it is important to note that all reflection models will predict an increase in reflected flux (with respect to the irradiating flux) above 10 keV, since the reflected flux is due to Compton backscattered emission, which causes the familiar `Compton hump' in the reflection spectrum.  Thus, whatever irradiating flux is assumed the reflection fraction should always increase between 10 -- 20 keV, as we find here.  The consequence of this is that the reflection models predict lags that should increase between 10 to 20 keV, opposite from what is observed.

This sharp contrast between the predicted reverberation lags and the observed lags above 8 keV leads us to conclude that reverberation alone cannot explain the lags in the lower kHz QPOs of \src, and that another physical mechanism is required to produce the observed lags.  A separate analysis of the lags in the lower kHz QPOs of \src\ by \citet{deavellar13} also finds that the lags decrease with increasing energy, again, at odds with our expectations from reverberation.  Moreover, \src\ is not the only neutron star LMXB to show lower kHz QPO lags that decrease with increasing energy -- both 4U~1636$-$36 \citep{deavellar13,deavellar16} and 4U~1728$-$34 \citep{peille15} also show the same behavior.  Thus, it appears that reverberation cannot be the physical mechanism behind the observed soft lags in lower kHz QPOs in those sources either.  \citet{peille15} also argue this based on the overall shape of the lag-energy spectrum in 4U~1728$-$34.  They discuss the models of \citet{kumar14} which utilize different mechanisms involved in thermal Comptonization as the source of the lags.  But, conclude that none of the simple models can satisfactorily explain all the spectral-timing features, including the shape of the covariance spectrum as well as the lags.  More recent work by \citet{kumar16} develops this Comptonization model further, and can qualitatively reproduce the observed energy-dependence of the lags. This model, however, predicts that the upper kHz QPO lags should also show the same energy-dependence.

The energy-dependence of the upper kHz QPO lags in \src, 4U 1636$-$36 \citep{deavellar13,deavellar16} and 4U~1728$-$34 \citep{peille15} show a markedly different behavior to those of the lower kHz QPO lags.  While the lower kHz QPO lags decrease with energy, the behavior of the upper kHz QPO lags in all three sources is much flatter, and in 4U~1728$-$34 at the highest frequencies it even increases with energy.  Our models here agree with the conclusions of \citet{peille15} that the general shape of those upper kHz QPO lags are much more consistent with a reverberation origin, at least qualitatively.  Thus, the lags in the lower and upper kHz QPOs may have a different physical origin.  In future work we will apply the methods developed here to fit the upper kHz QPO lags observed in 4U~1728$-$34, and determine whether they can be quantitatively explained by accretion disk reverberation.

\acknowledgements
EMC gratefully acknowledges support from the National Science Foundation through CAREER award number 1351222.

\bibliographystyle{apj}
\bibliography{apj-jour,felines}

\end{document}